# Probing the anisotropic behaviors of black phosphorus by transmission electron microscope, angular-dependent Raman spectra and electronic transports measurements

Wanglin Lu[1][♯], Xiaomeng Ma[2][♯], Zhen Fei[1], Jianguang Zhou[3], Zhiyong Zhang[2][*], Chuanhong Jin[1][*], Ze Zhang[1]

[1] *State Key Laboratory of Silicon Materials and School of Materials Science & Engineering, Zhejiang University, Hangzhou, Zhejiang 310027, China*

[2] *Key Laboratory for the Physics and Chemistry of Nanodevices and Department of Electronics, Peking University, Beijing 100871, China*

[3] *Research Center for Analytical Instrumentation, Institution of Cyber-Systems and Control, State Laboratory of Industrial Control Technology, Zhejiang University, Hangzhou, Zhejiang 310058, China*

In this study, we correlated the angular dependence of the Raman response of black phosphorus to its crystallographic orientation by using transmission electron microscopy and Raman spectroscopy. It was found that the intensity of the $A_g^2$ mode reached a maximum when the polarization direction of the incident light was parallel to the zigzag crystallographic orientation. Notably, it was further confirmed that the zigzag crystallographic direction exhibited superior conductance and carrier mobility. Because of the lattice extension along the armchair direction, an intensification of the anisotropic Raman response was observed. This work provides direct evidence of the correlation between anisotropic properties and crystallographic direction and represents a turning point in the discussion of the angular-dependent electronic properties of black phosphorus.

[♯] These authors contributed equally to this work

Author to whom correspondence should be addressed: zyzhang@pku.edu.cn; chhjin@zju.edu.cn.

Since the first report on the field-effect transistor (FET) device by the Zhang group, few- and single-layer black phosphorus (BP) has attracted intense research interest as an alternative single-element layered material to graphene.[1-7] The FET fabricated on a few-layer BP exhibits excellent electronic transport properties with an on/off ratio of $10^5$ and a carrier mobility of up to 1000 cm$^2$/V s.[1] According to theoretical calculations, the hole mobility for monolayer BP can reach as high as 10,000-26,000 cm$^2$/V s. The bandgap of the monolayer BP was calculated to be 1.51 eV,[8,9] with the gap dropping to 0.30 eV for its bulk counterpart; this compares favorably to the zero bandgap of graphene.[10] In microelectronics applications, Liu et al. successfully fabricated high-performance complementary metal-oxide-semiconductor inverters by combining p-type BP and n-type MoS$_2$.[11] For optoelectronic applications, Busceme et al. demonstrated the photovoltaic effect using the ambipolar transport behaviors of few-layer BP.[12,13] Additionally, the IBM group successfully fabricated high-performance photodetectors for multispectral and high-resolution imaging using few-layer BP.[14]

BP exhibits unique anisotropic properties,[15-21] including a highly anisotropic carrier mobility and angle-dependent optical response that arise from its puckering layered crystal structure;[22-24] these properties have been used to design novel optoelectronic and electronic devices.[13,25-30] For practical use of BP-based devices, it is important to develop a rapid and reliable method to directly correlate the crystallographic orientation (zigzag or armchair) with the anisotropic properties of the device. To solve this problem, several groups have developed experimental approaches using a combination of angle-dependent conductance and Raman spectroscopy.[31,32] A careful analysis of the observed polarization angle-dependent intensities of Raman active modes and their vibration directions enables the direct assignment of the zigzag and armchair orientations (in-plane) in BP. However, because these studies did not provide direct evidence for correlations of optical and electronic behaviors with crystallographic orientation, it is necessary to verify this correlation with other direct and accurate methods such as scanning tunneling microscopy (STM), transmission electron microscopy (TEM) and associated electron diffraction techniques.

In our experiments, polydimethylsiloxane (PDMS) was used for mechanical exfoliation and the following dry-transfer of few-layer BP samples at ambient environment.[33] To minimize the degradation, the humidity of the laboratory environment was controlled to be lower than 30%; For TEM observations, BP flakes were directly transferred onto an amorphous holey silicon nitride (SiN) TEM grids (Norcada Inc.), and then the suspended regions on the selected BP samples were characterized using an FEI Tecnai $G^2$-F20 TEM which was operated at an accelerating voltage of 200 kV; the electron dose was reduced to approximately 6.5 e/$Å^2$ s to minimize the radiation damage. Raman spectra were collected on a JY Horiba HR800 micro-Raman system with 532 nm laser excitation. The laser power was kept below 0.1mW to avoid laser-induced damage. A 100× objective lens was used for all Raman spectra measurements. The sample was rotated in a custom-built rotation stage with a step of 5°-10°. Six pairs of electrodes along the directions from 0° to 360° with an inter-electrode step of 30° were fabricated to measure the electronic conductance. Evaporated electrodes along zigzag or armchair direction were used as transistors for extracting the field effect mobility.

An optical image of an as-exfoliated BP flake with a thickness of 8.3 nm is presented in Fig. 1(a). The region of interest (marked by the red rectangle in Fig. 1(a)) can be precisely traced during TEM observations as shown in Fig. 1(b). As illustrated in Fig. 1(c), high-resolution TEM and selected-area electron diffraction (SAED) characterizations were carried out to identify the crystalline orientation; SAED patterns were recorded along the [010] zone axis in the marked BP flake region. The measured lattice spacings are 3.37±0.01Å and 4.42±0.01 Å in the (100) direction and in the horizontal direction for the (001) lattice, respectively. These results match well with the lattice orientation along zigzag and armchair directions, respectively, as demonstrated by the model shown in the inset of Fig. 1(c). The SAED pattern shows that the BP flake had excellent crystallinity; this was critical for achieving credible optical and electronic characterizations.

After the TEM determination of crystalline orientation, the same BP flake was used for the angular dependence Raman spectrum characterization. A parallel-polarized configuration was first used, with the polarization of the collecting analyzer parallel to that of the incident laser beam (532 nm wavelength). For ease of analysis, we used electron field vectors $e_i$ and $e_s$ to separately denote the polarization of the incident and collected scattered light with respect to the BP flake (shown in Fig. 1(a)); θ was the rotation angle of the BP zigzag crystallographic direction with respect to the polarization direction of incident beam. Angle-dependent intensities of $A_g^2$, $B_{2g}$ and $A_g^1$ modes were presented only from 0 ° to 180 ° (as shown in Figure 2), considering the two-fold rotational symmetry of BP along the incident direction of the laser. It was found that the Raman intensities showed periodic oscillation as a function of the θ angle. The $A_g^2$ mode achieved its maximum (local maximum) intensity when the rotation angle was 0 °(90 °); the $B_g^2$ mode exhibited a 90 ° period and achieved its minimum intensity at both 0 ° and 90 °. Combined with the above TEM results on crystallographic orientation, our results indicate that $A_g$ modes reach maximum intensity when the polarization of the incident laser is parallel to the zigzag crystallographic direction of BP, which was in opposite to the previously reported results.[31,32] Because the accumulated intensity of a Raman mode is determined by the Raman cross-section, which is proportional to $|e_i \times R \times e_s|^2$, the intensities are related to the Raman tensor R and the scattering geometry in polarization configuration e.[34,35] According to Eq. (S1), the Raman tensor parameter c is larger than the parameter a under 532 nm laser excitation, consistent with the recently obtained theoretical result.[36] To qualitatively understand the superiority of the zigzag orientation for the optical response, we may consider that Raman intensity may be principally influenced by the photon-electron excitation and electron-phonon interaction processes. Because the excitation can be explicitly influenced by the light polarization angle with respect to the BP crystallographic direction, a much stronger absorption of light may arise for light polarized along the zigzag direction.[37,38] This zigzag-dominated process for the optical response may contribute to the maximum intensities of the $A_g$ modes. Considering that the armchair direction is inherently pliable, a higher emission efficiency may occur when the laser polarization and the vibration direction of $A_g^2$ mode are in parallel. This may explain the appearance of the local maximum when light polarization occurs parallel to the armchair direction.

To more conveniently identify the crystallographic orientations, we measured the angular dependence Raman spectra without the analyzer in the collection path. As shown in Fig. 2(d)-(f), all observed vibration modes show similar oscillation behaviors except for the $B_{2g}$ mode; this is because the excitation of this mode is not prohibited when θ is 45 ° or 135 °.[34] Based on this experimental fact, we can unambiguously identify the crystalline orientation with this simplified configuration. Because we have confirmed the relationship between the crystallographic orientation and angular dependence Raman response with the combination of TEM and Raman measurements, we can now correctly identify the crystalline orientation by angle-resolved Raman spectroscopy using a green laser; this led us to a deeper consideration of the relationship between linear dichroism and the crystallographic orientation.

To further corroborate our experimental findings, we also carried out multiple-electrode device characterizations and measured the angle-dependent conductance of BP sheets to confirm the relationship between conductance and crystallographic orientation. The angle-dependent Raman spectra were first recorded to identify the crystallographic orientation on a BP flake that was subsequently used in device fabrication. Six pairs of electrodes were fabricated on BP sheets along the defined directions spaced by 30 ° between 0 ° to 360 ° relative to the crystallographic orientation, as shown in Fig. 3(a). The normalized intensity plot of $A_g^2$ mode is presented in Fig. 3(b) and shows that the maximum value is obtained for the 90 ° angle corresponding to the zigzag crystallographic direction, whereas a local maximum value is obtained at 0 ° corresponding to the armchair direction. The normalized angular dependence conductance was measured by applying a fixed electric field across each pair of diagonally positioned electrodes. The measured anisotropic conductance results indicated that the maximum conductance $\sigma_{max}$ was obtained along 90 °, corresponding to the zigzag direction, whereas the minimum conductance $\sigma_{min}$ was obtained along 0 °, corresponding to the armchair direction with the ratio $\sigma_{max}/\sigma_{min}$ of approximately 2. Hence, we could deduce that BP exhibits superior (inferior) conductance along the zigzag (armchair) direction.

Anisotropic field effect mobility was measured in few-layer BP devices fabricated on several neighboring flakes with the same thickness (as shown in Fig. S1(b)). Transistor channels were deliberately constructed along the zigzag and armchair direction separately as shown in Fig. 4(a) (pre-determined by angle-dependent Raman spectra measurement), and the measurements were conducted in vacuum conditions at room temperature by applying a voltage of -1 V as the source-drain bias. We extracted the corresponding mobility using the expression $\mu_{FE} = \left(\frac{dI_{ds}}{dV_{bg}}\right) \times [L/(W C_g V_{ds})]$, where $L$ and $W$ are, respectively, the length and width of the channel, $C_g$ is the gate capacitance, and $V_{ds}$(1V) is the drain-source bias. The calculated field effect mobility along the zigzag and armchair crystallographic directions are approximately 110 cm$^2$/V s and 78 cm$^2$/V s, respectively. Based on the mobility statistics obtained from several BP samples (as listed in Table S1 in ref 34), we found that the field effect mobility along the zigzag direction is approximately 1.4 times higher than that along the armchair direction. The difference in the carrier density may be ascribed to the fact that the anisotropic mobility ratio is slightly smaller than the anisotropic conductance ratio. As reported recently, the carrier mobility along the armchair direction should be higher than that along the zigzag direction.[22,32] We believe that this finding is a turning point for the investigation of anisotropic properties of BP sheets and that these results should motivate further theoretical analysis.

We repeated the measurements on various samples with different thicknesses down to three layers. A typical atomic force microscopy (AFM) image of such a thin BP flake is shown in Fig. 5(a). Because the BP interlayer spacing is 5.3 Å, the height of the thin film (2.1 nm) produced here indicated a tri-layer film. As a typical vibration mode, the $A_g^2$ mode maintained the same oscillation period. However, the ratio between the maximum intensity and the local maximum intensity rose drastically by a factor of almost 2 for the tri-layer film compared with the thick samples (Fig. 5(b)). SAED pattern and HRTEM image of the tri-layer BP sheet are presented in Fig. 5(c)-(d). Based on the SAED data and the Fourier transform of the HRTEM image, we calculated that the lattice parameter along the armchair direction increased by 0.06 Å, whereas the lattice parameter along the zigzag direction remained almost unchanged; this is consistent with the previously reported computational results.[8] The thickness-dependent Raman intensity can be explained by the changes in Raman tensor elements in R influenced by the variation of electron-phonon dispersion[38] that typically arises from the changes in atomic structure.[7,19,39,40] Hence, we believe that the stretching or shrinking of the lattice parameter significantly influences the dielectric constant and electron-phonon dispersion, which then modulate the a and c Raman tensor element values. Although linear dichroism may be affected by the changes in incident laser light wavelength,[41] it is reasonable to believe that the micro-structure evolution induced by the variation in the thickness of BP changes the polarization behavior of Raman intensities at a fixed wavelength. The $A_g^2$ mode could therefore be regarded as a reliable fingerprint for the identification of the crystallographic direction in BP forms varying from bulk to tri-layer.

In summary, we have carried out an independent investigation of the correlation between the anisotropic optical and electronic properties of BP and its crystallographic orientation. Based on the correlation between the crystalline orientation determined by TEM characterization and the measured

Raman behaviors and conductance, we proposed that the zigzag crystallographic direction exhibits superior optical response and electronic conductance. Dramatic anisotropic structure evolution and a resultant rise in the $A_g^2$ intensity oscillation were observed when the thickness of the BP flake was reduced from bulk to tri-layer. The results of this study may provide further understanding of linear dichroism and electron-phonon dispersion and help build high-performance devices utilizing the superior properties of the BP zigzag crystallographic direction.

The authors would like would like to thank Prof. Wencai Ren for kindly providing high-quality BP crystals, Prof. Wei Ji for fruitful discussions, and the Center of Electron Microscopy of Zhejiang University for the access to TEM facilities. This work was financially supported by the National Basic Research Program of China (2014CB932500, 2015CB921000), the National Science Foundation of China (51222202, 51410305074, 51472215 and 61427901); the Program for Innovative Research Team in University of Ministry of Education of China (IRT13037) and the Fundamental Research Funds for the Central Universities (2014XZZX003-07). J.G.Z. was supported by the National Key Technology Support Program (2012BAB19B00) and the National Key Scientific Instrument and Equipment Development Project (2013YQ470781).

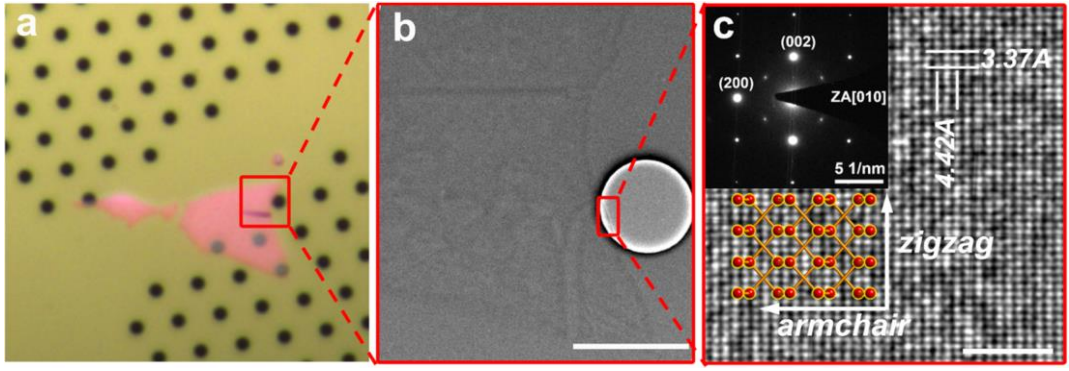

FIG. 1. (a) An optical image of an 8.3-nm-thick BP flake transferred onto a SiN TEM window. θ is the angle between zigzag crystallographic direction and polarization direction of incident laser. The scale bar is 5 μm. (b) The low-magnified TEM image of the selected region of black phosphorus (as marked with red rectangles in Fig. 1(a)). The scale bar is 2 μm. (c) HRTEM image of the selected suspended region as marked in (b). The scale bar is 2 nm.

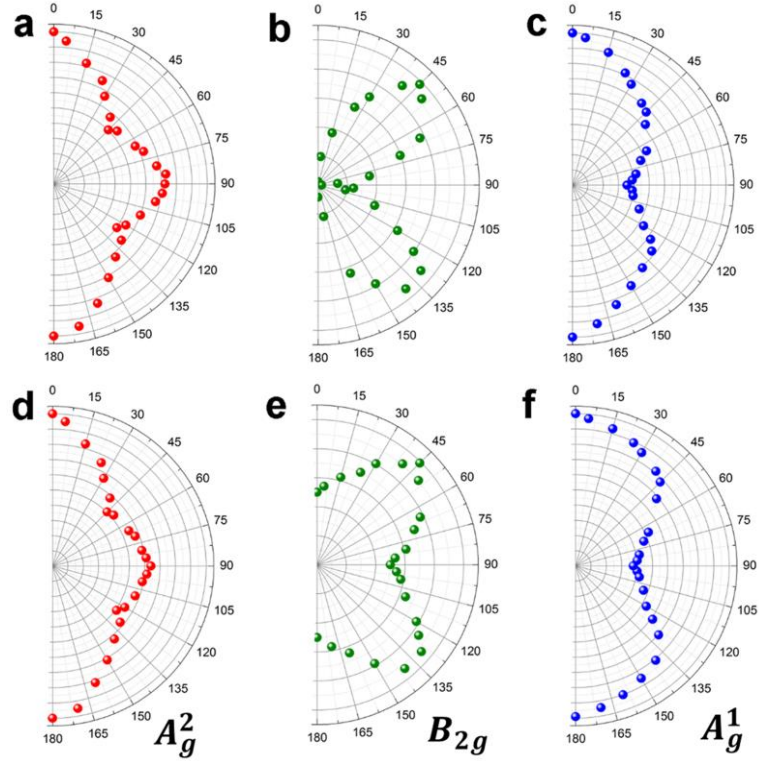

FIG. 2. (a), (b) and (c) Polar plots of the angle-dependent normalized Raman intensities of $A_g^2$, $B_{2g}$ and $A_g^1$ modes. Here, an analyzer was placed in parallel to the polarization of incident light in the collection path. (d), (e) and (f) Polar plots of the angle-resolved normalized Raman intensities of $A_g^2$, $B_{2g}$ and $A_g^1$ modes, without an analyzer in the collection path.

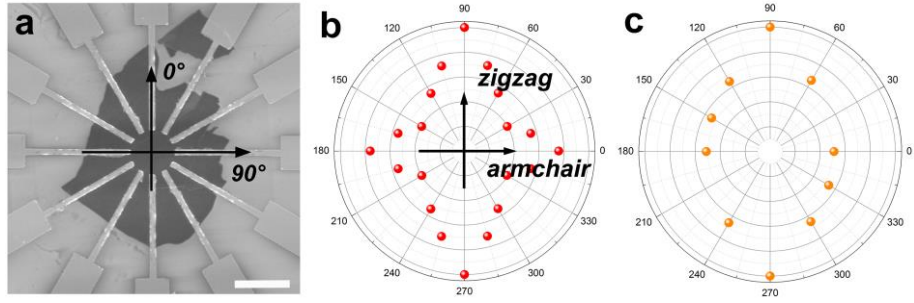

FIG. 3. (a) An SEM image showing a few-layer BP flake based device with six pairs of electrodes spaced by 30°. The scale bar is 5 μm. (b) Polar plots of the normalized intensity of $A_g^2$ mode as a function of rotation angle θ (defined in Fig. 1(a)). The intensity maximum and local maximum correspond to the zigzag and armchair crystallographic direction, respectively. (c) Polar plots of the normalized conductance as a function of angle θ.

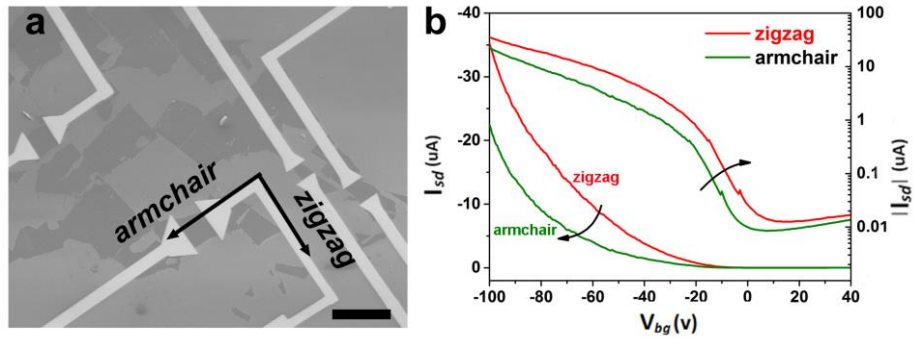

FIG. 4. (a) An SEM image of transistors with the channels fabricated along zigzag or armchair crystallographic directions. The scale bar is 20μm. (b) Room-temperature anisotropic transfer characteristic along zigzag (red online) and armchair (green online) directions, with source-drain bias of -1 V.

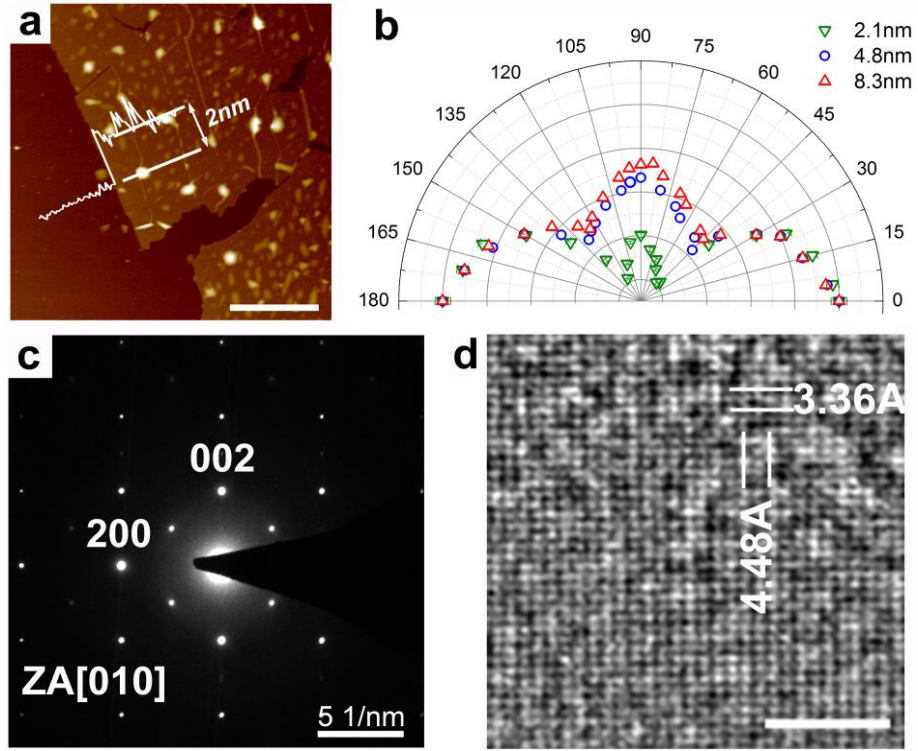

FIG. 5. (a) An AFM image of a few-layer BP sheet with a height of 2.0 nm. (b) Polar plots of the angle-resolved normalized intensities of $A_g^2$ mode recorded on three typical flakes with different thickness. (c), (d) show the SAED and HRTEM images recorded on the tri-layer BP flake.

# Supporting information

# Probing the anisotropic behaviors in black phosphorus by TEM, angular-dependent Raman spectra and electronic transports measurements

Expressions for angular dependence of Raman modes:

The intensities of Raman modes are determined by Raman tensor and scattering geometry. Raman tensors of $A_g$ and $B_{2g}$ modes can be expressed as:

$$R(A_g) = \begin{pmatrix} a & 0 & 0 \\ 0 & b & 0 \\ 0 & 0 & c \end{pmatrix} \qquad R(B_{2g}) = \begin{pmatrix} 0 & 0 & d \\ 0 & 0 & 0 \\ d & 0 & 0 \end{pmatrix}$$

Since we just consider parallel configuration, polarization vectors of light can be expressed:

$$e_i = e_s = (\sin\theta \quad 0 \quad \cos\theta)$$

Raman scattering intensity S for parallel polarization configuration is defined as

$$S \propto |e_i \times R \times e_s|^2$$

So

$$S(A_g) \propto (a\sin^2\theta + c\cos^2\theta)^2 \qquad (1)$$

$S(A_g)$ achieves maximum when $\theta$ is 0° and 90° respectively, hence,

$$S(A_g)_{max} = c^2; \; S(A_g)_{min} = a^2$$

$$S(B_{2g}) \propto (2d\sin\theta\cos\theta)^2 \qquad (2)$$

When we withdraw the analyzer in the light path while the incident laser beam is still polarized at the same direction, the Raman scattering intensity S is defined as

$$S \propto |e_i \times R|^2$$

$$S(A_g) \propto (a\sin\theta + c\cos\theta)^2 \qquad (3)$$

$$S(B_{2g}) \propto d^2(\sin\theta + \cos\theta)^2 \qquad (4)$$

For this configuration, $B_{2g}$ will be permissible at 0 ° or 90 °, and we have

$$\frac{S_{max}}{S_{min}} = \frac{2d^2}{d^2} = 2 \qquad (5)$$

So, with this configuration, $B_{2g}$ mode is not prohibited.

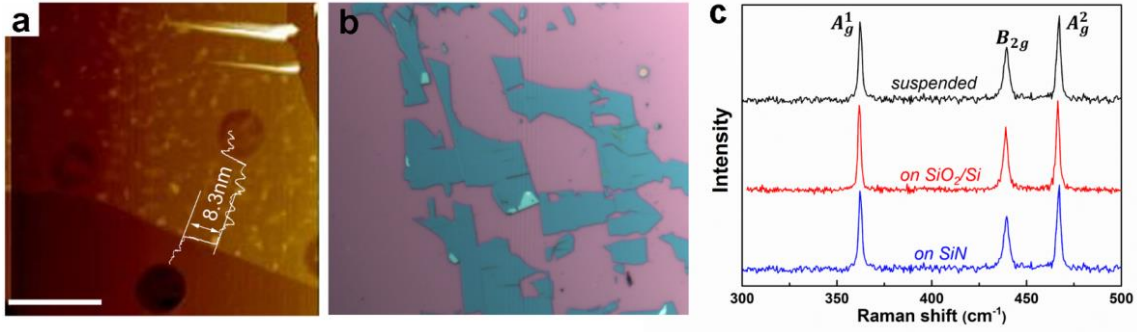

FIG. S1. (a) An AFM image of a thick BP film (about 8.3 nm, over 15 layers). (b) A typical low-mag optical image showing a large area BP flakes exhibiting the same optical contrast. These flakes are believed to be identical considering they were exfoliated from the same bulk area. (c) Typical Raman spectra of thick BP flakes on different substrates: SiN, SiO$_2$/Si and substrate-free, respectively. No noticeable differences in Raman frequencies and intensity ratios are observed.

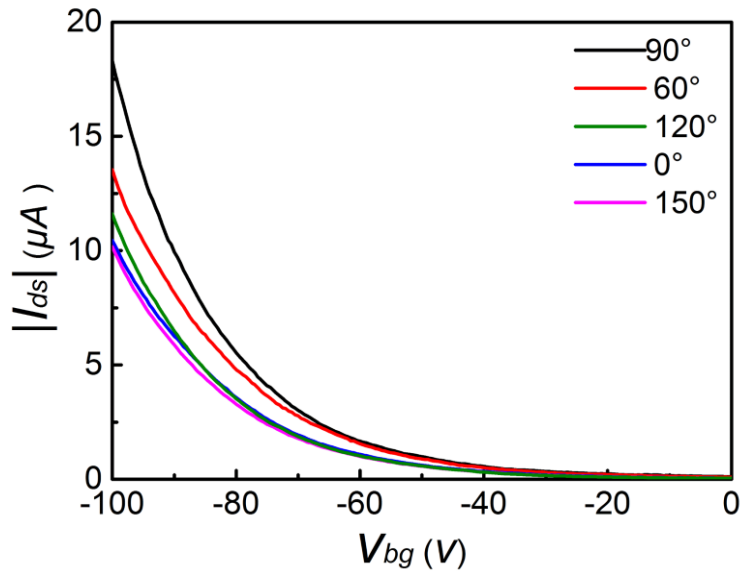

FIG. S2. The transfer characteristic of the multiple-electrode device shown in Fig. 3(a).

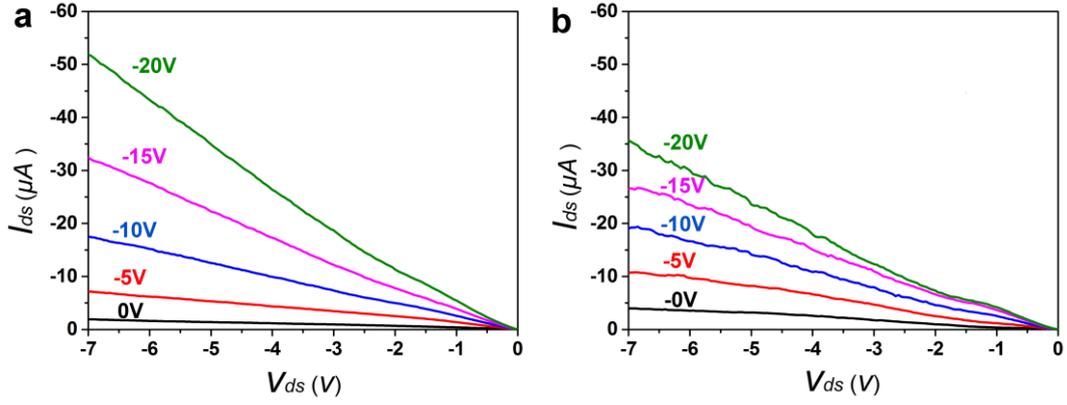

FIG. S3. (a) The output characteristics of transistors with the channel fabricated along zigzag direction. The channel length and width are 7.5μm and 6.1μm respectively. (b) The output characteristics of transistors with channel along armchair direction. The channel length and width are 8.4μm and 7.7μm respectively.

TABLE I Statistics of carrier mobility for several samples.

|  | 1-Zig | 1-Arm | 2-Zig | 2-Arm | 3-Zig | 3-Arm |
|---|---|---|---|---|---|---|
| L/W | 7.5/6.1 | 8.4/7.7 | 5.3/3.3 | 9.3/9.4 | 3.7/3.8 | 5/4.9 |
| Mobility ($cm^2$/V s) | 109.6 | 78.3 | 136.1 | 98.8 | 259 | 194.2 |
| Zig/Arm ratio | 1.3997446 | | 1.37753 | | 1.333677 | |

1-Zig (1-Arm) is sample 1 with channel fabricated along zigzag direction.

Zig/Arm ratio means the mobility along zigzag direction to the mobility along armchair direction.